# Evidence of electron interaction with an unidentified bosonic mode in superconductor $CsCa_2Fe_4As_4F_2$

Peng Li[1], Sen Liao[1], Zhicheng Wang[2], Huaxun Li[3], Shiwu Su[1], Jiakang Zhang[1], Ziyuan Chen[1], Zhicheng Jiang[4,5], Zhengtai Liu[6], Lexian Yang[7], Linwei Huai[8], Junfeng He[8], Shengtao Cui[4], Zhe Sun[4], Yajun Yan[1], Guanghan Cao[3], Dawei Shen[4,5], Juan Jiang[1,†], Donglai Feng[1, 4,5,9†]

[1]Shool of Emerging Technology, University of Science and Technology of China, Hefei 230026, China

[2]Key Laboratory of Quantum Materials and Devices of Ministry of Education, School of Physics, Southeast University, Nanjing 211189, China

[3]School of Physics, Zhejiang University, Hangzhou 310058, China

[4]National Synchrotron Radiation Laboratory, University of Science and Technology of China, Hefei, 230026, China

[5]School of Nuclear Science and Technology, University of Science and Technology of China, Hefei, 230026, China

[6]Shanghai Synchrotron Radiation Facility, Shanghai Advanced Research Institute, Chinese Academy of Sciences, Shanghai 201210, China

[7]State Key Laboratory of Low Dimensional Quantum Physics, Department of Physics, Tsinghua University, Beijing 100084, China

[8]CAS Key Laboratory of Strongly-coupled Quantum Matter Physics, University of Science and Technology of China, Hefei, Anhui 230026, China

[9]New Cornerstone Science Laboratory, University of Science and Technology of China, Hefei, 230026, China

Emails: jjiangcindy@ustc.edu.cn; dlfeng@ustc.edu.cn

**Abstract:**

The kink structure in band dispersion usually refers to a certain electron-boson interaction, which is crucial in understanding the pairing in unconventional superconductors. Here we report the evidence of the observation of a kink structure in Fe-based superconductor $CsCa_2Fe_4As_4F_2$ using angle-resolved photoemission spectroscopy. The kink shows an orbital selective and momentum dependent behavior, which is located at 15 meV below Fermi level along the $\Gamma - M$ direction at the band with $d_{xz}$ orbital character and vanishes when approaching the $\Gamma - X$ direction, correlated with a slight decrease of the superconducting gap. Most importantly, this kink structure disappears when the superconducting gap closes, indicating that the corresponding bosonic mode (~ $9 \pm 1$ meV) is closely related to superconductivity. However, the origin of this mode remains unidentified, since it cannot be related to phonons or the spin resonance mode (~ 15 meV) observed by inelastic neutron scattering. The behavior of this mode is rather unique and challenges our present understanding of the superconducting paring mechanism of the bilayer FeAs-based superconductors.

## Introduction

Electron-boson coupling is the cornerstone for pairing in superconductors. In Bardeen-Cooper-

Schrieffer theory for conventional superconductors, phonons are identified as the pairing glue. While for unconventional superconductors, such as cuprates and Fe-based superconductors, antiferromagnetic (AFM) magnons or spin excitations are often considered as the potential pairing bosonic mode. Particularly, spin resonance mode observed by inelastic neutron scattering (INS) correlates with superconductivity. For example, it emerges below the superconducting transition temperature ($T_C$), its energy ($E_R$) correlates with $T_C$ and its momentum connects different parts of the Fermi surface [1-12]. The resonance energy $E_R$ approximately follows an empirical ratio $E_R/2\Delta \sim 0.64$, where $\Delta$ is the superconducting gap [3,13,14]. The observation of a smaller $E_R$ compared to $2\Delta$ provides evidence for the existence of a spin exciton. On the other hand, numerous experiments reported the evidence of electron-phonon coupling in these materials [15-25], and other proposals of pairing glue such as orbital fluctuations [26-28], add more complexity to the investigation of unconventional superconductivity.

Recently, a stochiometric Fe-based superconductor, $ACa_2Fe_4As_4F_2$ (A=K, Rb, Cs) [29], has attracted attention due to their multiple-layer structures and a high $T_C$ of around 30 K. However, its pairing symmetry remains controversial. Several muon-spin relaxation (μsR) measurements [30-32] suggested the possible nodal superconducting gaps, while other methods indicated the nodeless gap feature [33-38]. Interestingly, INS experiments reported resonance peaks with energies even higher than $2\Delta$ which cannot be understood by the spin exciton picture [13,39]. This sets them apart from other Fe-based superconductors, and the presence of such atypical resonance peaks appears to challenge the AFM spin fluctuation scenario, sparking renewed interest in their pairing mechanisms.

In this Letter, we present a comprehensive angle-resolved photoemission spectroscopy (ARPES) investigation of the low-energy electronic structure of $CsCa_2Fe_4As_4F_2$ ($T_C\sim$ 29 K). Our findings reveal a pronounced kink structure that is strongly correlated with superconductivity. It appears at a binding energy ($E_B$) of 15 meV along the $\Gamma - M$ direction exclusively in the $\alpha$ band ($d_{xz}$), and disappears above the superconducting gap closing temperature. The kink exhibits strong angular dependent behavior, and diminishes in the $\Gamma - X$ direction, giving a dramatic change in the electron-boson coupling strength ($\lambda_{e-b}$). Additionally, we observe that the anisotropic superconducting gap of the $\alpha$ band closely follows the anisotropic behavior of this kink. Based on the size of the corresponding superconducting gap and kink energy, we propose a bosonic mode with an energy of around 9 meV which is highly related to superconductivity in this bilayer system. However, unlike the kink observed in $Ba_{0.6}K_{0.4}Fe_2As_2$ [40], the kink in $CsCa_2Fe_4As_4F_2$ *cannot* be directly attributed to the reported spin resonance mode and its origin remains mysterious.

**Results and Discussion**

The crystal structure of $CsCa_2Fe_4As_4F_2$, depicted in Fig. 1(a), can be regarded as a composite of CaFeAsF and $KFe_2As_2$. It exhibits a total doping level of 0.25 hole/Fe due to the self-doping effect [29]. Figures 1(b-d) show the basic electronic structure of the sample. The measured Fermi surface using 50 eV photons at 10 K is presented in Fig. 1(b). Around the Γ point, there are three hole pockets (α, β and γ) while four hot spot-like pockets (δ) surround the M point, more details could be found in Supplementary Fig. 1 and Supplementary Fig. 2. One should notice that as a bilayer system, there should be two sets of Fe orbitals, and we neglect this in our discussion hereafter due to the fact that the interlayer coupling is weak in this system and the splitting of these bands is beyond our experimental resolution [35]. The quasi two-dimensional electronic structure of this compound is confirmed by $k_z$-dependent measurements in Supplementary Fig.3, which is consistent with previous transport measurements where there was huge resistivity anisotropy of $\rho_c(T)/\rho_{ab}(T)$ up to $10^3$ [41]. In addition to the primary bands, there exists a weak large pocket around the M point, as indicated by the white dashed curves. This is believed to be the folded $\gamma$ band from the Γ point (black solid circle) due to surface reconstruction, similar to other alkali metal intercalated Fe-based superconductors [42,43], indicating most likely the measured samples are with Cs terminated surface. Both linear-horizontal (LH) and linear-vertical (LV) polarizations are utilized to detect the orbital characters of these bands, as shown in Figs. 1(b) and (c). The LV polarization exhibits a stronger intensity of the $\alpha$ band, indicating its $d_{xz}$ orbital nature [44]. The $\beta$ and $\gamma$ bands around Γ are formed by $d_{yz}$ and $d_{xy}$ orbitals, respectively, while the $\delta$ and $\varepsilon$ bands at the M point consist of $d_{yz}$ and $d_{xy}$ orbitals, respectively. Notably, a distinct kink structure is observed in the $\alpha$ band well below $T_C$ as depicted in Fig. 1(e), located at 15 meV below the Fermi level. This kink anomaly is distinct from a Bogoliubov bending band since the energy positions differs (Supplementary Fig. 4) and it is also confirmed by our scanning tunneling microscopy (STM) result in Supplementary Fig. 5. More details about the kink will be discussed in the later.

Figure 2 displays the superconducting gaps and their angular dependence across different bands. The symmetrized spectra along the Γ − M direction measured at 7 K shown in Figs. 2(a) and 2(b) reveal a distinct gap opening at $\alpha$, $\beta$ and $\delta$ bands. It should be noted that the $\gamma$ band may exhibit a tiny gap beyond energy resolution. The superconducting gap values (Δ) for the $\alpha$, $\beta$ and $\delta$ bands are 6 meV, 6 meV, and 5.5 meV, respectively [Fig. 2(c) and Supplementary Fig. 6], determined from the fitting of the symmetrized energy distribution curves (EDCs) in a BCS-based phenomenological model [45]. The temperature evolution of these bands can be found in Supplementary Fig. 7. Figures 2(d)-(e) present momentum distributions of the symmetrized EDCs for the $\alpha$ and β bands. The extracted gap values reveal a slight anisotropy in the $\alpha$ band, with the gap size gradually decreasing from 6 meV in the Γ − M direction to 4 meV in the Γ − X direction as shown in Fig. 2(f). However, the $\beta$ band is nearly isotropic, with a constant gap size of 6 meV as shown in Fig. 2(g). More details can be found in Supplementary Fig. 8.

The kink feature is exclusively observed at the $\alpha$ band and strongly coupled with superconducting phase (Fig. 3). The criterion of the existence of kink is that both the real (ReΣ) and imaginary part (ImΣ) of the self-energy (Σ) appear anomalies around the kink energy. When the superconducting gap closes, both the raw spectrum (the lower two panels of Fig. 3(a)) and the extracted peak positions (Fig. 3(d)) indicate an absence of this kink, where we use multiple Lorentzian peaks to present the fitted bands as shown in Supplementary Fig. 9. In the light of the extracted momentum distribution curves (MDCs) for the $\alpha$ band at 7 K (Fig. 3(b)), we have identified the kink position located at 15 meV below $E_F$ highlighted by the blue curve, where there is a dramatic change in slope. The self-energy analysis of the $\alpha$ band is shown in Fig. 3(c). Both the real and imaginary parts of the self-energy exhibit anomalies around this energy, and the ReΣ displays a relatively broad feature (ranging from 6 meV to 35 meV) with a maximum at 15 meV. Such a broad energy span is peculiar and intriguing (the data can be repeated on another sample in Supplementary Fig. 10). The Kramers-Kronig (KK) transformation analysis is shown in Supplementary Fig. 11, which presents that both ReΣ and ImΣ match well with each other under the KK transformation, indicating the self-consistency of the extracted self-energies. The strength of electron-boson coupling ($\lambda_{e-b}$) of the kink can be determined by analyzing the slope change in both bare and renormalized band dispersions, where $\lambda_{e-b}$ is defined as:

$$\lambda_{e-b} = \frac{(\frac{d\varepsilon}{dk})_{bare}}{(\frac{d\varepsilon}{dk})_{renormalized}} - 1 \qquad (1)$$

We use the curve at 100K as the bare band, and the energy range of the renormalized band lies between the gap energy and the kink energy. The extracted $\lambda_{e-b}$ values gradually decrease with the increasing temperature, and eventually disappears above the superconducting gap closing temperature, as shown in Fig. 3(e). Details of the kink structure evolution with temperature can be found in Supplementary Fig. 12-14. One could notice that the kink seems to persist slightly above the transition temperature 29 K and vanishes around 35 K, which is possible due to the existence of pseudogap in this material as reported by several other experiments[46,47].

We further analyze the angular distribution of the kink in the $\alpha$ band well below $T_C$ in Fig. 4. Figures. 4(a) and 4(d) depict the representative high symmetry cuts along the $\Gamma - X$ and $\Gamma - M$ directions, respectively (more details can be found in Supplementary Fig. 15). The angle offset from the $\Gamma - M$ direction is defined as $\theta$. The kink structure is present along the $\Gamma - M$ direction, while it is absent along the $\Gamma - X$ direction. The self-energy analysis also reveals discernible distinctions between these two directions, as illustrated in Figs. 4(b) and 3(c). In contrast to the observable anomalous signals around 15 meV in the $\Gamma - M$ direction, no significant variation is observed in both Re$\Sigma$ and Im$\Sigma$ in the $\Gamma - X$ direction. The peak positions of the $\alpha$ band are extracted step by step from the $\Gamma - M$ direction to the $\Gamma - X$ direction with a 5-degree increment, as summarized in Fig. 4(c) with momentum shifts. The dashed straight lines serve as guides for the kink positions at each curve and the determination of the kink position is in Supplementary Fig. 16. Figure 4(e) shows that both the kink energy position $E_{kink}$ and the corresponding coupling strength $\lambda_{e-b}$ exhibit remarkable decreasing trends as $\theta$ increases. The kink vanishes beyond 25°, accompanied by a slight energy shift from 15 meV at 0° to 13 meV at 25°. Notably, the coupling strength also dramatically decreases towards zero at the edge of 25°. In spectroscopic experiments, electron-boson interactions may give rise to low energy anomalies near the Fermi level, appearing at energies of $\Delta + \Omega$ in principle, where $\Delta$ represents the superconducting gap and $\Omega$ denotes the corresponding bosonic energy. We extract the boson energies at various angles by subtracting the $\Delta_\alpha$ from $E_{kink}$. As shown in Fig. 4(f), the extracted boson energies are located at a constant value of $9 \pm 1$ meV. Therefore, it is reasonable to infer that the distribution of the electron-boson coupling correlates with the anisotropic superconducting gap of the $\alpha$ band [48].

A common origin of such a kink in band dispersion is electron-phonon coupling, which renormalizes the Fermi velocity around the phonon energy $\Omega$[49]. For ACa$_2$Fe$_4$As$_4$F$_2$ (A=K, Rb, Cs), numerous phonon modes have been detected in various experiments [34,46,50,51], and these modes were observed in both the normal and superconducting phases. In our ARPES experiment, however, the absence of kink behavior above $T_C$ exclude phonon as a candidate. The 9 meV bosonic mode is directly associated with a mode that should only emerge when the superconducting condensation is ready, and several other experiments have also reported a potentially superconducting correlated bosonic mode in ACa$_2$Fe$_4$As$_4$F$_2$ [36,38,52].

The spin resonance has been discovered to emerge in the superconducting states of cuprates and iron-based superconductors [2,4,6,7,11,53-60]. While the correlation between kink structures and the corresponding spin resonance modes have been widely reported in cuprates [9,61-63], in which these kinks are located at the energy $E_{kink} = \Delta + E_R$ [63], only a few Fe-based superconductors, such as Ba$_{0.6}$K$_{0.4}$Fe$_2$As$_2$ [40] and (Sr/Ba)$_{1-x}$K$_x$Fe$_2$As$_2$ [64], were found to exhibit this correspondence. Notably, the kink discovered in CsCa$_2$Fe$_4$As$_4$F$_2$ which presents a broad distribution of Re$\Sigma$ cannot be accounted for by the sharp spin resonance mode at 15 meV observed by the neutron scattering [13][39], as the ARPES determined kink corresponds to a boson energy of $9 \pm 1$ meV. Therefore, we exclude the possibility of the reported spin resonance mode as being the physical origin of the kink. However, it is noteworthy that the Re$\Sigma$ of the $\alpha$ band presents a broad feature in energy where the spin resonance mode may exist within this broad spectrum, but with a limited impact.

These bilayer superconductors (ACa$_2$Fe$_4$As$_4$F$_2$) are quite unique among all iron-based superconductors. On one hand, it shows giant anisotropy between the in-plane and out-of-plane resistivity which is different from most iron pnictides and mimic cuprates[41]. On the other hand, the spin resonance modes possess a unique downward dispersion that is similar to cuprates in previous INS report, and its energy exceeds the limit of 2Δ[13], challenging the conventional understanding of the resonance modes. Due to the similarities to cuprates, one possible candidate for causing such a broad ReΣ in energy may be the existence of paramagnon, which is broad in energy and only observed in numerous cuprates by the resonant inelastic soft X-ray scattering but not by neutron scattering [65]. Another possible mode that emerges in superconducting state is Josephson plasmon, which only exists in the multiple-layered superconductors [66-69]. Further optical measurements on thick samples are needed to study the Josephson plasmon. If true, the correlation between momentum dependent gap and $\lambda_{e-b}$ may suggest that the electron-Josephson plasmon can further facilitate pairing in addition to other correlation effects. One could notice that, similar to spin resonance mode and electron-magnon coupling in multi orbital system [40,70], this mode either originated from paramagnon or Josephson plasmon also has a strong orbital selectivity here. Additionally, the extracted boson energy closely aligns with the spin gap energy (~ 10 meV) of this compound [13], suggesting that the spin gap might be also the possible candidate similar to the case in the overdoped Bi$_2$Sr$_2$CaCu$_2$O$_{8+\delta}$ [71]. Considering that the strong superconductivity correlated kink structures are rather rare in other iron-based superconductors, such a remarkable kink in band dispersion is a unique feature observed for the bilayer system.

In conclusion, our ARPES study of CsCa$_2$Fe$_4$As$_4$F$_2$ presents a unique case that deviates from many other studied Fe-based superconductors. We reveal an anomalous kink in CsCa$_2$Fe$_4$As$_4$F$_2$ that exists just at the $\alpha$ band, which is most likely caused by electron interacting with certain bosons. This kink exhibits strong coupling with superconductivity and correlates with the superconducting gap of the $\alpha$ band. However, it cannot be attributed to phonons and the spin resonance found by INS. Our results may provide a counterexample that challenges the critical role of spin resonance in Fe-based superconductors. Finally, we believe that both the unique spin resonance mode in INS and the unique bosonic mode in our ARPES measurements can uncover some anomalous but critical aspects. The identity of this unique bosonic mode calls for further study which could be crucial to understand the superconducting mechanism in Fe-based superconductors.

**Methods**

High quality CsCa$_2$Fe$_4$As$_4$F$_2$ single crystals were synthesized by the solid-state reaction method [29,39]. Synchrotron-ARPES measurements were performed at beamline BL03U of Shanghai Synchrotron Radiation Facility (SSRF) and BL13U of National Synchrotron Radiation Laboratory (NSRL) in China . The overall energy resolution for the gap measurement was set to be better than 6 meV at 23 eV photon energy and the angular resolution is ~ 0.2 degree for the gap and kink measurements. The crystals were cleaved *in-situ* and measured with a base pressure better than $6 \times 10^{-11}$ Torr. Lab-based laser ARPES measurements were carried out using 6.999 eV light source at University of Science and Technology of China. The overall energy resolution was set to be better than 2 meV the angular resolution is ~ 0.2 degree for the gap and kink measurements. All the data presented in this paper were taken within a few hours after cleavage ensuring the results were not affected by aging effect.

**Data Availability**

All data needed to evaluate the conclusion in the paper are present in the paper and the Supplementary information. All raw data generated during the current study are available from the corresponding author upon request.

coupling in the overdoped $Bi_2Sr_2CaCu_2O_{8+\delta}$. *Nat. Commun*. **11**, 569 (2020).

**Acknowledgements**

This work is supported by the National Natural Science Foundation of China (Grant No. 12174362 and No. 11888101, No. 92065202), the Innovation Program for Quantum Science and Technology (Grant No. 2021ZD0302800, No. 2021ZD0302802), the New Cornerstone Science Foundation. Part of this research used Beamline 03U of the Shanghai Synchrotron Radiation Facility, which is supported by ME2 project under contract no. 11227902 from National Natural Science Foundation of China. The authors would like to express their gratitude for the insightful discussions with H. Q. Luo, X. Y. Lu. S. S. Qin, K. Jiang, and J. P. Hu.

**Author Contributions Statement**

P. L. and J. J. conceived the experiments. P. L., S. L., S. W. S. and J.J. carried out ARPES measurements with the synchrotron assistance from Z. C. L., Z. T. L. and D. W. S. at SSRF, S. T. C. and Z. S. at NSRL and laser-ARPES assistance from L. X. Y., L. W. H. and J. F. H.. Z. C. W, H. X. Li and G. H. C. synthesized single crystals. J. K. Z., Z. Y. C. and Y. J. Y. conducted the STM experiments. P. L., J. J. and D. L. F. wrote the manuscript. All authors contributed to the scientific planning and discussions.

**Competing Interests Statement**

The authors declare no competing interests.

**Figure Legends**

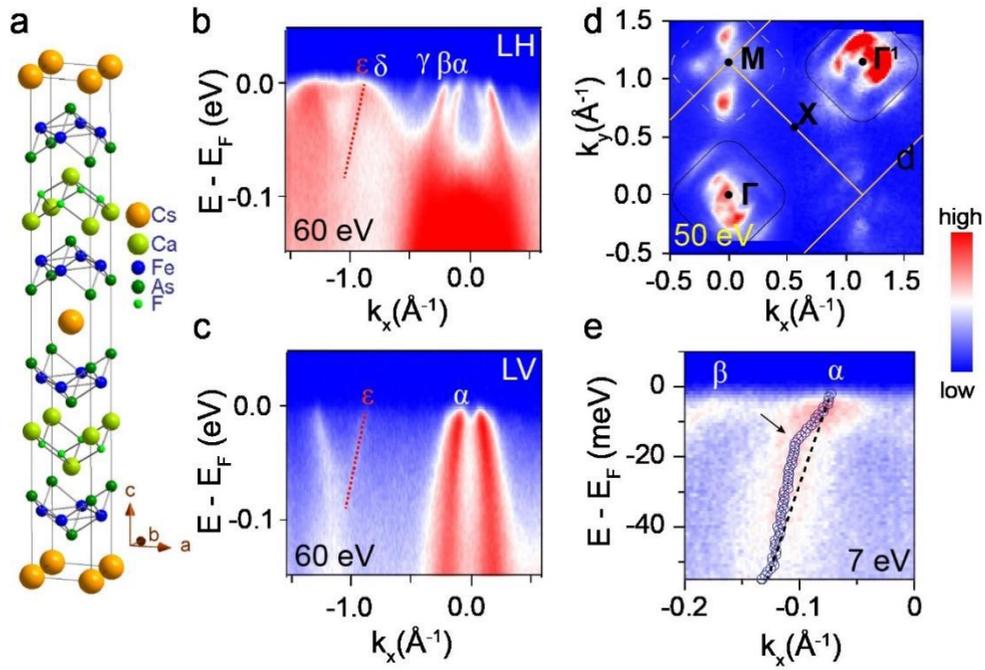

**FIG. 1.** Basic crystal and electronic band structure of CsCa$_2$Fe$_4$As$_4$F$_2$. (a) The crystal structure of CsCa$_2$Fe$_4$As$_4$F$_2$. (b) and (c) ARPES spectra along the Γ − M direction by using both LH and LV polarized 60 eV photons. The bands are denoted by α, β, γ, δ and ε. (d) Fermi surface map obtained at 10 K using 50 eV photons, superposed with 2Fe-Brillouin zone (BZ) boundary (yellow solid lines) and high symmetry points (black dots). (e) Kink structure on the α band measured with 7 eV laser. The black arrow indicates the kink position and the black dashed line indicates the bare band in normal state.

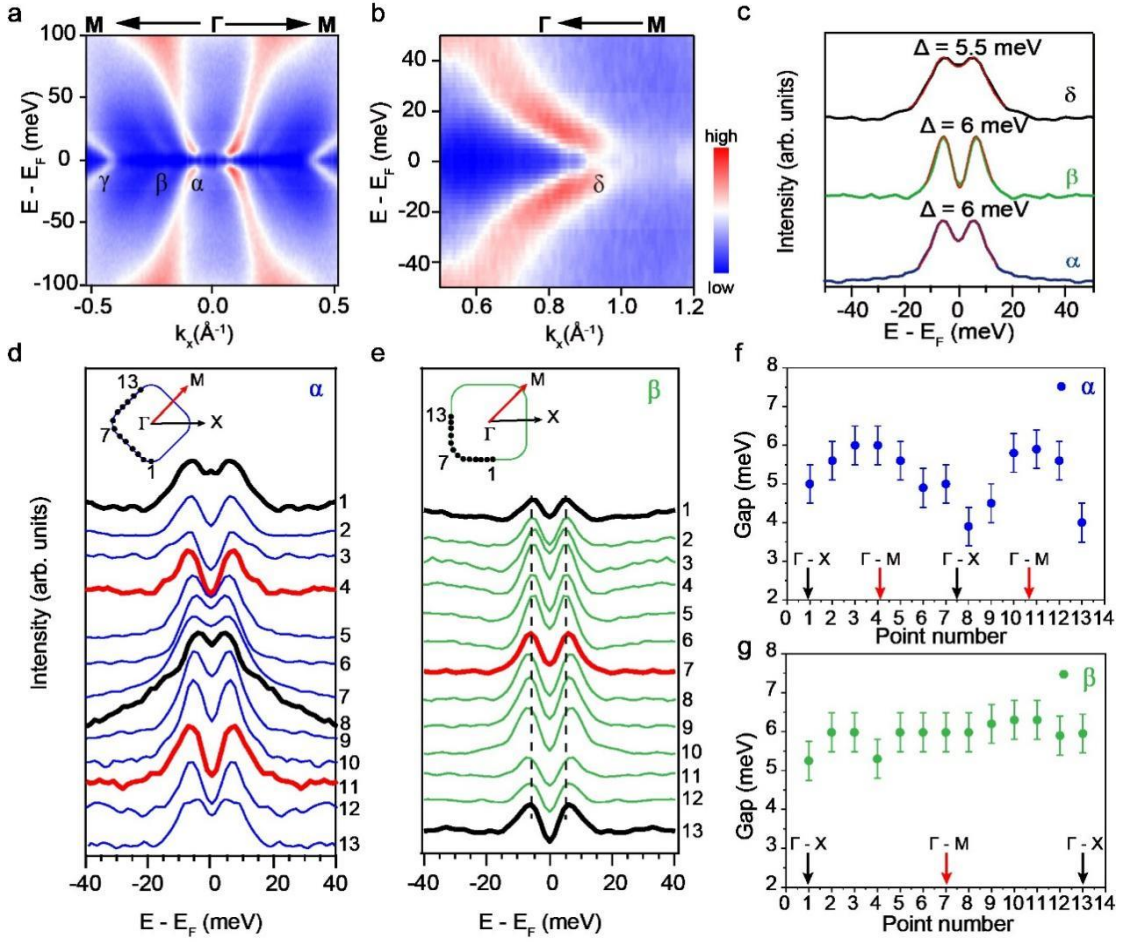

**FIG. 2.** Multiband superconducting gaps and their momentum distributions. (a) Symmetrized ARPES spectrum along the $\Gamma - M$ direction measured at 7 K with 23 eV photons, where $\alpha$, $\beta$ and $\gamma$ denote distinct bands. The color bar shows the ARPES spectra intensity. (b) Symmetrized ARPES spectrum of the $\delta$ band measured near the M point along the $\Gamma - M$ direction measured at 7 K with 50 eV photons. (c) Symmetrized EDCs of the $\alpha$ (7 eV, blue curve), $\beta$ (7 eV, green curve) and $\delta$ (50 eV, black curve) bands. The red lines are fits using phenomenological model. Their superconducting gap sizes are annotated above. (d), (e) Symmetrized EDCs of the points noted in the insets of the $\alpha$, $\beta$ pockets, respectively. (f) Distribution of the in-plane superconducting gap of $\alpha$ (upper panel), $\beta$ (lower panel) pockets, respectively. The red and black arrows indicate the $\Gamma - M$ and $\Gamma - X$ directions, respectively. The error bars reflect the uncertainty in determining the gap sizes.

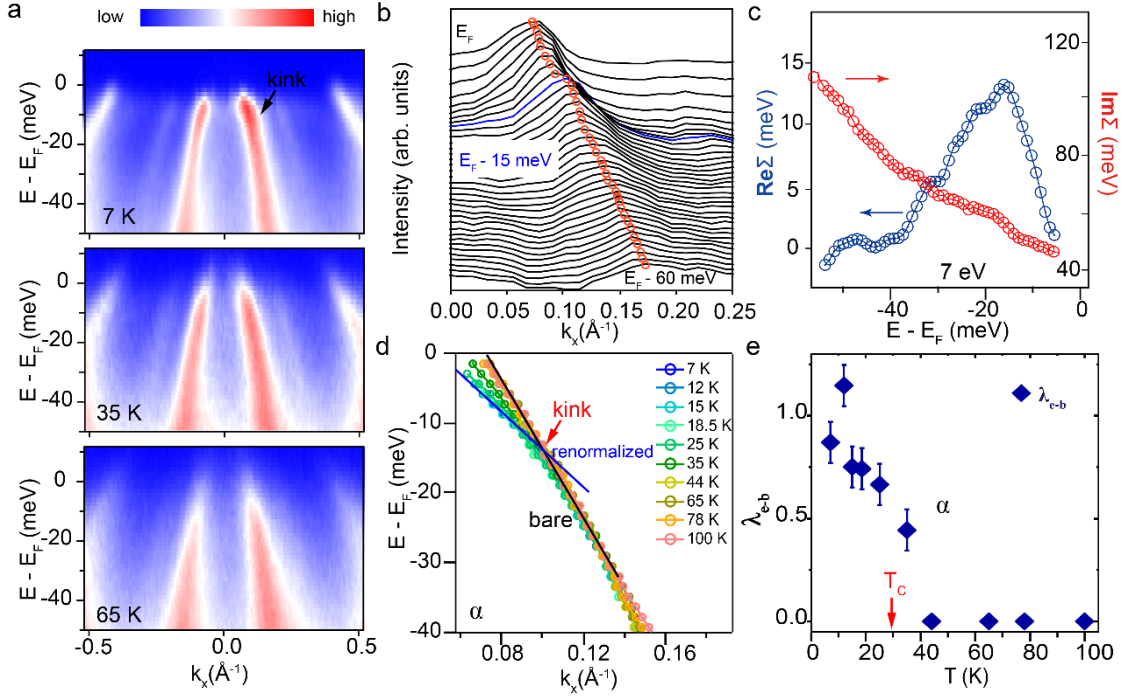

**FIG. 3.** Kink in the α band and its temperature evolution properties. (a) ARPES Spectra measured along the Γ − M direction at 7 K, 35 K and 65 K. The presence of a kink in the α band is indicated by the black arrow at 7 K. The color bar indicates the photoemission intensity. (b) Corresponding MDCs of the α band measured at 7K in panel (a), where peak positions are marked with circles and the kink position is highlighted in blue. (c) The self-energy analysis of the kink reveals anomalies in both ReΣ (blue circle curve) and ImΣ (red circle curve) at an energy of approximately 15 meV below the Fermi level. The color bar shows the ARPES spectra intensity. (d) The peak positions of the α band, extracted from MDCs at various temperatures. (e) The corresponding electron-boson coupling strength $\lambda_{e-b}$ as a function of temperature. The red arrow marks the $T_C$. The error bars reflect the uncertainty in determining the coupling strength.

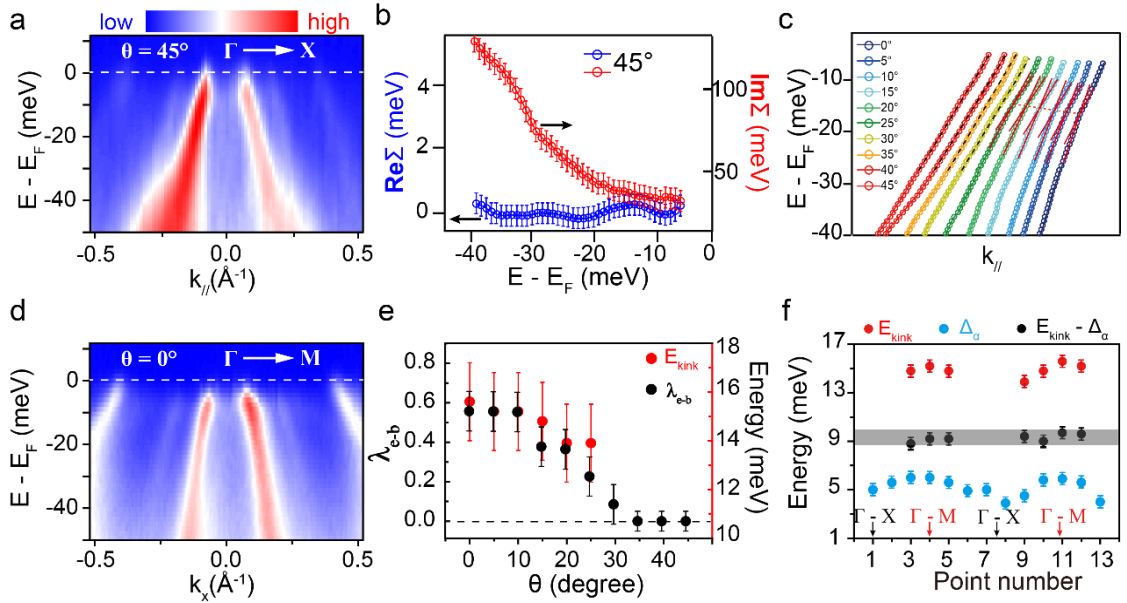

**Fig 4.** Momentum dependent behaviors of the kink in the α band. (a), (d) The ARPES spectra measured along the Γ − X direction with θ = 45° and the Γ − M direction with θ = 0°, respectively. The color bar

indicates the photoemission intensity. (b) Re$\Sigma$ and Im$\Sigma$ of the α band for $\theta = 45°$ which show negligible anomaly. (c) In-plane dispersions of the α band extracted from their corresponding MDC curves presented at $5°$ intervals from the $\Gamma - M$ to the $\Gamma - X$ directions, which are stacked with momentum offsets for a better illustration. (e) The kink energy positions $E_{kink}$ and electron-boson coupling strength $\lambda_{e-b}$, extracted from (c), as a function of $\theta$ angles. They both exhibit suppression upon approaching the $\Gamma - X$ direction. (f) A summary of the $E_{kink}$ (red dots) and the superconducting gap $\Delta_\alpha$ (light blue dots) in the α band. The kink-corresponding bosonic mode energies are presented as the black dots with a near constant energy of $9 \pm 1$ meV. The error bars reflect the uncertainty in determining the gap sizes and coupling strength.